\documentclass[prd, onecolumn, notitlepage, nofootinbib,showkeys, amsmath, amssymb, aps, superscriptaddress]{revtex4-2}

\usepackage{graphicx}% Include figure files
\usepackage{dcolumn}% Align table columns on decimal point
\usepackage{bm}% bold math
\usepackage{xcolor}
\usepackage{hyperref}% add hypertext capabilities
\hypersetup{
    colorlinks = true, 
    linkcolor = black,
    citecolor = black,
    urlcolor = teal,
}
\usepackage{lipsum}
\usepackage{mathrsfs}

\usepackage[T1]{fontenc}
\usepackage[utf8]{inputenc}

\usepackage{tabularx}
\newcolumntype{Y}{>{\centering\arraybackslash}X}
\usepackage{multirow}
\usepackage{nicefrac}

\usepackage{ulem}

\newcommand{\dder}{\textup{ d}}
\newcommand{\Dder}{\textup{ D}}

\begin{document}

\title{Geodesic dynamics in brane--de Sitter wormholes}

\author{W. S. Kl\"{e}n}
\email{wayner.klen@gmail.com}
\affiliation{Instituto de F{\'{\i}}sica, Universidade de S\~{a}o Paulo. \\
 Rua do Mat{\~ a}o 1371, CEP 05508-090, S{\~ a}o Paulo - SP, Brazil.
}
\affiliation{Instituto Tecnol{\'o}gico de Aeron{\'a}utica. \\
Pra{\c c}a Marechal Eduardo Gomes 50, CEP 12228-900, S{\~ a}o Jos{\'e} dos Campos - SP, Brazil.
}

\author{R. C. Barboza}
%\email{rodbarboza9502@usp.br}
\email{r.barboza9502@gmail.com}
\affiliation{Escola de Artes, Ci{\^ e}ncias e Humanidades, Universidade de S\~{a}o Paulo. \\
Avenida Arlindo B{\' e}ttio 1000, CEP 03828-000,  S{\~ a}o Paulo - SP, Brazil.
}

\author{C. Molina}
\email{cmolina@usp.br}
\affiliation{Escola de Artes, Ci{\^ e}ncias e Humanidades, Universidade de S\~{a}o Paulo. \\
Avenida Arlindo B{\' e}ttio 1000, CEP 03828-000,  S{\~ a}o Paulo - SP, Brazil.
}

\begin{abstract}

We present a dynamical analysis of the null and timelike geodesics around an asymptotically de Sitter wormhole in a Randall-Sundrum brane. In this framework, the wormhole throat is interpreted both as a photon sphere and as a fixed point of the associated dynamical system. The stability of this structure is evaluated using Lyapunov and Jacobi criteria with consistent results. A Bogdanov-Takens bifurcation is observed in the null-geodesic dynamics, highlighting critical changes in the behavior of light around the wormhole. Explicit solutions are derived for geodesics near the throat, providing insight into the optical appearance of the wormhole shadow. These results show qualitatively similar behavior for null and timelike orbits, suggesting universal features of geodesic dynamics in brane--de Sitter wormholes.

\end{abstract}


\maketitle

\section{Introduction}
\label{sec:Intro}

General relativity is a central theory in our understanding of the Universe, providing a framework for the analysis of a wide range of astronomical and cosmological phenomena and has been extensively tested with increasing accuracy. Notable recent observations include the direct detection of gravitational waves by LIGO and Virgo in 2015 \cite{Abbott2016} and the first direct image of a black hole detected by the Event Horizon Telescope in 2019 \cite{EHT2019}.
Despite its enormous success, attempts have been made to extend general relativity. One such approach is the braneworld models, designed as concrete implementations of string theories in a cosmological framework. Among the wide variety of brane models, we highlight the Randall-Sundrum-type scenarios \cite{RD1999-1,RD1999-2}, where a four-dimensional brane describing our Universe is immersed in a five-dimensional bulk. In this context, solutions describing compact objects such as black holes and wormholes have been proposed. Relevant examples of this research can be found in \cite{Germani:2001,CFM2002,Lemos:2003,Bronnikov1,Bronnikov2,Abdalla2006,Lobo:2007,Molina:2010yu,Molina:2011mc,Neves2012,Neves2012it,Molina:2013mwa,Parsaei:2015,Molina:2016tkr,Ghosh:2021}.

Currently, we are primarily interested in wormhole solutions. These structures are compact spacetimes with nontrivially topological interiors and topologically simple boundaries \cite{Visser1995}. They can be thought of as connections between distant parts of the Universe, while still being compatible with usual local physics.
Wormholes have been the subject of extensive research since the seminal papers of Morris, Thorne, and Yurtsever \cite{Morris1988,Morris_Thorne_Yurtsever}. Although the actual existence of wormholes remains speculative, the very fact that these objects are consistent solutions compatible with general relativity raises profound questions about causality and time travel \cite{Morris_Thorne_Yurtsever,Konoplya,Ovgun}. Research on this topic continues to provide new insights, ideas, and challenges to general relativity and its extensions \cite{Konoplya:2005,Klen2020,Neves:2021dqx,Frizo:2023,Neves:2024zwi}.

Trajectories of light in the vicinity of ultracompact objects (UCOs) can be viewed as a gravitational signature of the system, providing valuable information about the background geometry \cite{Klen2020}. Of particular interest are the photon spheres of UCOs. These structures are closed paths of light in the neighborhood of the central object, strongly depending on the characteristics of the UCO's spacetime. From an observational point of view, photon spheres play a key role in defining the shadows generated by UCOs. Beyond their observational significance, photon spheres are closely related to perturbative analyses of spacetime. The correspondence between photon spheres and quasinormal modes has been extensively explored \cite{Cardoso2009,Chen2022,Anacleto2021}. These studies reveal a compelling link between gravitational waves and geodesic properties \cite{Cardoso2009,Konoplya2017,Morgan2009}.

The connection between geodesics and dynamical systems has been the subject of considerable research. For example, the study of geodesics in the presence of black holes has provided new information about the behavior of these objects and has helped to further our understanding of their properties \cite{Chiba2017, Mohaddese, Mohaddese-2}. In addition, the investigation of geodesic dynamics in the expanding Universe has provided new insights into the evolution of the Universe as a whole \cite{Conroy2014,cunningham2017}. 

Dynamical systems sensitive to parameter variations can undergo structural changes in their phase space topology. These changes can be manifested by the creation or annihilation of equilibrium points or by modifications to their stability properties. In the dynamical systems literature, such qualitative transformations of the state-space topology induced by parameter variations are called bifurcations \cite{kuznetsov-book,kuznetsov2005}. Bifurcations serve as fundamental mechanisms for behavioral shifts in dynamical systems, generating preferred regions of phase space or allowing transitions to complex dynamics, including quasiperiodic and chaotic trajectories.
In this context, critical phenomena attract significant attention. For instance, bifurcations in solutions of Einstein's field equations have been extensively studied, from standard Friedmann-Lema\^{\i}tre-Robertson-Walker cosmologies \cite{Kohli2018, Kokarev2009} to modified theories of gravity \cite{Antoniou2018,Doneva2018,Azim2020}. 
In recent work, these concepts have been extended to the analysis of the interactions of dark matter and dark energy \cite{Aydiner2025}.
Alternative bifurcation mechanisms in scalar theories have also been identified, leading to the formation of hairs in charged and rotating black holes \cite{Herdeiro2018,Herdeiro2014}. Homoclinic bifurcations in the dynamics of geodesics, which can lead to chaotic behavior in the system, have been studied in the context of perturbed Schwarzschild black holes \cite{Bombelli1992}. Furthermore, Bogdanov-Takens bifurcations have been explored in braneworld scenarios, revealing potential gravitational signatures \cite{Klen2020}. 

The main goal of the present work is to apply the formalism of dynamical systems to the framework involving geodesics around wormholes. A family of geometries describing de Sitter wormholes in asymptotically de Sitter Randall-Sundrum brane models was derived in \cite{Neves2012}. We will revisit the solutions presented in \cite{Neves2012} using the language of dynamical systems. In a sense, the proposal here is to continue the work developed in \cite{Klen2020}, where an extensive dynamical analysis of the null geodesics in the spacetimes proposed by Casadio, Fabbri, and Mazzacurati was carried out. The wormhole throat will be dynamically characterized, and its stability is evaluated by Lyapunov and Jacobi criteria. Connections between geometric features of the wormhole spacetime and the dynamics of null geodesics in these backgrounds will be investigated.

The structure of the paper is presented as follows. In Sec.~\ref{sec:Wormholes}, the wormhole spacetimes explored in this paper are revisited, and the coordinate systems of interest are defined.
In Sec.~\ref{sec:Potential}, an effective potential is constructed for null and timelike geodesics, based on the quasilocal coordinate $u$. The main properties of this potential are established.
The geodesic dynamics is analyzed in Sec.~\ref{sec:dynamics}. 
For this purpose, an effective two-dimensional dynamical system is constructed, and in this formalism, the wormhole throat is characterized as a photon sphere and as a fixed point.
The stability of the photon sphere is determined by Lyapunov and Jacobi criteria.
Bifurcations and other special aspects of the null-geodesic dynamics are discussed, with an examination of the optical appearance of the wormhole as seen by a static observer.
The timelike geodesic dynamics, and its relation to the null case counterpart, are discussed.
Final comments are presented in Sec.~\ref{sec:Final}.
In this paper, we use signature $(-,+,+,+)$ and geometric units with $G_{4D}=c=1$, where $G_{4D}$ is the effective four-dimensional gravitational constant.

\section{\label{sec:Wormholes}de Sitter Wormholes in the brane}

\subsection{Asymptotic de Sitter braneworld solutions}

The physical scenario considered in the present work is an asymptotically de Sitter wormhole in a braneworld setup \cite{Neves2012,Neves:2024zwi}. The spacetime of interest is interpreted as a four-dimensional brane immersed in a five-dimensional bulk. The solution was derived from the treatment proposed by Maeda, Sasaki, and Shiromizu \cite{MSS2000}. In this formalism, the field equations in the brane, as a generalization of the four-dimensional Einstein's field equations, can be written as
\begin{equation}
R_{\mu\nu}-\frac{1}{2} R g_{\mu\nu}=-\Lambda g_{\mu\nu}-\mathcal{E}_{\mu\nu} \, ,
\label{eq_projetada}
\end{equation}
with vacuum assumed in the brane. 
The four-dimensional metric, Ricci tensor, and Ricci scalar are represented by $g_{\mu\nu}$, $R_{\mu\nu}$, and $R$ respectively. 
The tensor $\mathcal{E}_{\mu \nu}$, the projection of the five-dimensional Weyl tensor in the brane, furnishes the bulk's gravitational influence over the brane. 
The effective cosmological constant on the brane is given by $\Lambda$.

Using the fact that $\mathcal{E}_{\mu \nu}$ is traceless, a suitable combination of the effective field equations~\eqref{eq_projetada} can be obtained,
\begin{equation}
R=4\Lambda\, ,
\label{Ricci_scalar}
\end{equation}
where $R$ is the four-dimensional Ricci scalar. 
Imposing staticity and spherical symmetry, the four-dimensional line element can be written as
\begin{equation}
\textup{d} s^2 = -A(r) \textup{d} t^2 + \frac{\textup{d} r^2}{B(r)} + r^2 \left(\textup{d} \theta^2 + \sin^2 \theta  \textup{d} \phi^2 \right) \, ,
\label{Metric_ansatz}
\end{equation}
with the chart $\{ t,r, \theta, \phi\}$. The coordinate $r$ is the ``areal radius'' and $\theta$ and $\phi$ are the usual angle coordinates associated with spherical symmetry. 
The combination~\eqref{Ricci_scalar} can be seen as a constraint between the metric functions $A(r)$ and $B(r)$ \cite{Neves2012}, namely
\begin{equation}
2(1-B)-r^2 B \left[ \frac{A''}{A}-\frac{(A')^2}{2A^2}+\frac{A'B'}{2AB} + \frac{2}{r}\left( \frac{A'}{A}+\frac{B'}{B}\right) \right] = 4\Lambda r^2 \, ,
\label{Dif_constraint}
\end{equation}
where ($'$) denotes differentiation with respect to $r$.

The most general solution of the constraint~\eqref{Dif_constraint} with the condition $A(r) = B(r)$ is given by
\begin{equation}
A_{0} (r) = B_{0} (r) = 1 - \frac{2M}{r} + \frac{q}{r^2} - \frac{\Lambda}{3} r^2 \, .
\label{eq_RNdS}
\end{equation}
The Reissner-Nordstr\"{o}m--de Sitter spacetime has the same form as the metric defined by $A(r) = A_{0} (r)$ and \linebreak $B(r) = B_{0} (r)$ in Eq.~\eqref{Metric_ansatz}. However, the constant $q$ in the braneworld setup is interpreted as a ``tidal charge'' that is not related to the electric properties of the compact object described by the four-dimensional geometry. Instead, it is associated with the properties of the brane within a bulk, which means that there is no restriction on the sign of $q$. The functions $A_{0} (r)$ and $B_{0} (r)$ of the metric define the base solution.

In the nonextreme regime (the focus of this paper), $q < q_{\textrm{ext}}$ and $0 < \Lambda < \Lambda_{\textrm{ext}}$, where $q_{\textrm{ext}}$ and $\Lambda_{\textrm{ext}}$ are bounds on the tidal charge and the cosmological constant, respectively. 
In this case, $A_{0}(r)$ has four real roots, denoted $r_n$, $r_-$, $r_+$ and $r_c$, with $r_n < 0 < r_- < r_+ < r_c$.

The solution of interest in the present work can be considered ``close'' to the base solution, in the sense that it is a continuous deformation of the electrovacuum solution~\eqref{eq_RNdS} that satisfies the constraint~\eqref{Dif_constraint} \cite{Molina:2010yu,Molina:2011mc,Neves2012,Neves2012it,Molina:2013mwa,Molina:2016tkr},
\begin{equation}
A(r) = A_{0} (r) = \frac{\Lambda}{3r^2}(r_c -r)(r-r_+)(r-r_-)(r-r_n) \, ,
\label{eq:A(r)}
\end{equation}
\begin{equation}
B(r) = A_{0}(r) \left[ 1+(C-1) P(r) \right] \, ,
\label{eq:B(r)}
\end{equation}
where the function $P(r)$ is 
\begin{equation}
P(r) \equiv 
\left(
\frac{r_{c} - r_{0}}{r - r_{0}}
\right)^{c_{0}}
\left(
\frac{r_{c} - r_{0-}}{r - r_{0-}}
\right)^{c_{0-}}
\left(
\frac{r_{c} - r_{0--}}{r - r_{0--}}
\right)^{c_{0--}}
\left(
\frac{r_{c} - r_{0n}}{r - r_{0n}}
\right)^{c_{0n}}
\, .
\label{def-P}
\end{equation}
The constants $r_{0}$, $r_{0-}$, $r_{0--}$ and $r_{0n}$ are the (simple and real) zeros of the function $h(r)$, 
\begin{equation}
h(r) \equiv  \frac{r A'_{0}(r)}{2} + 2 A_{0}(r)
 = 2 - \frac{3M}{r} + \frac{q}{r^{2}} - \Lambda r^2 \, ,
\end{equation}
with $r_{0n} < 0 < r_{0--} < r_{0-} < r_{0}$.
The coefficients $c_{0}$, $c_{0-}$, $c_{0--}$ and $c_{0n}$ are given by
\begin{eqnarray}
c_{0} & = & \frac{2}{\Lambda} \, \frac{r_{0}\left(2\Lambda r_{0}^{2} - 1\right)}{\left(r_{0} - r_{0-}\right)\left(r_{0} - r_{0--}\right)\left(r_{0} - r_{0n}\right)} \, , \\
c_{0-} & = & - \frac{2}{\Lambda} \, \frac{r_{0-}\left(2\Lambda r_{0-}^{2} - 1\right)}{\left(r_{0} - r_{0-}\right)\left(r_{0-} - r_{0--}\right)\left(r_{0-} - r_{0n}\right)} \, , \\
c_{0--} & = & \frac{2}{\Lambda} \, \frac{r_{0--}\left(2\Lambda r_{0--}^{2} - 1\right)}{\left(r_{0} - r_{0--}\right)\left(r_{0-} - r_{0--}\right)\left(r_{0--} - r_{0n}\right)} \, , \\
c_{0n} & = & -\frac{2}{\Lambda} \, \frac{r_{0n}\left(2\Lambda r_{0n}^{2} - 1\right)}{\left(r_{0--} - r_{0n}\right)\left(r_{0} - r_{0n}\right)\left(r_{0-} - r_{0n}\right)} \, .
\end{eqnarray}
The interpretation and bounds for the parameter $C$ will be discussed in the following.%
\footnote{In \cite{Neves2012}, the definition of the parameter $C$ is different from the one employed in the present work. Hence, the slight difference form for the metric function in Eq.~\eqref{eq:B(r)}.}

\subsection{Extending the spacetime}
\label{sec:wormhole} 

From the explicit solution for the metric in Eqs.~\eqref{eq:A(r)} and \eqref{eq:B(r)}, the global properties of the brane can be discussed. These properties strongly depend on the value of $C$. This constant can be interpreted as a deformation parameter related to the properties of the brane.

For instance, it is straightforward to verify that
\begin{equation}
A(r) \sim B(r) \sim 
1 - \frac{\Lambda}{3}r^2  + O \left( \frac{1}{r} \right) \, .
\end{equation}
That is, the spacetime is asymptotically de Sitter for any value of $C$. 

A key question concerns the range of the parameter $C$ and its implications for the geometry of the system. The value of $C$ determines the nature of the spacetime described by Eqs.~\eqref{eq:A(r)} and \eqref{eq:B(r)}. Specifically, for $C>1$ the geometry has a singularity enclosed by a cosmological horizon. For $C=1$, the base solution is recovered, with a singularity enclosed by a Cauchy horizon. The case of interest for this work is when
\begin{equation}
0 < C < 1 \, .
\label{eq:range-C}
\end{equation}
In the range~\eqref{eq:range-C}, the function $B(r)$ has a simple zero $r_{\text{thr}}$ such that $A(r_{\text{thr}})\ne 0 \,$. Considering the relation between $r_{+}$, $r_{0}$, $r_{\textrm{thr}}$ and $r_{c}$, the following inequality is satisfied:
\begin{equation}
r_+ < r_0 < r_{\textrm{thr}} < r_{c} \, .
\label{inequality}
\end{equation}
The metric functions $A(r)$ and $B(r)$ are positive and analytic for $r_{\textrm{thr}} < r < r_{c}$, and the coordinate system $(t,r,\theta,\phi)$ is valid only in this domain. The maximal extension of the geometry has the structure of a wormhole, with a throat at $r = r_{\textrm{thr}}$, covered by a cosmological horizon at $r=r_c$. There is no event horizon or singularity in this spacetime \cite{Neves2012}.

A coordinate system in which the throat is directly represented is based on the so-called quasilocal radial coordinate $u$, defined as \cite{Bronnikov1, Bronnikov2}
\begin{equation}
\frac{\textup{d} u}{\textup{d} r} = \sqrt{\frac{A(r)}{B(r)}} \, .
\label{def-u}
\end{equation}
The function $u(r)$ is strictly positive and crescent in the interval $r_{\textrm{thr}} < r < r_{c}$, with a well-defined inverse function $r(u)$. 
Using the chart $(t,u,\theta,\phi)$, the line element is written as
\begin{equation}
\textup{d} s^2 = -\mathcal{A}(u) \textup{d} t^2 + \frac{\textup{d} u^2}{\mathcal{A} (u)}  + r(u)^2 \left(\textup{d} \theta^2 +\sin^2 \theta \ \textup{d} \phi^2 \right) \, ,
\label{metric_u}
\end{equation}
where 
\begin{equation}
\mathcal{A}(u) \equiv A(r(u)) \, .
\end{equation}

With a suitable choice of integration constant for Eq.~\eqref{def-u}, the throat $r=r_{\textrm{thr}}$ is mapped into $u=0$ and the cosmological constant $r=r_c$ is mapped into $u=u_c \equiv u(r_c)$, with $0 < u_c < \infty$. 
The extension ``beyond $r_{\textrm{thr}}$'' can be implemented as
\begin{equation}
r_{\textrm{thr}} < r < r_c \,\, \longrightarrow \,\,  -u_c < u < u_c \, .
\end{equation}
The extension ``beyond $r_{c}$'' can be done with the usual Eddington-Finkelstein coordinates. After the maximal extension, the spacetime describes a symmetric and transversable wormhole, with a throat at $u=0$. The two sections of the wormhole (with $u<0$ and $u>0$) are bounded by cosmological horizons, and the geometry is asymptotically de Sitter.

Analytic expressions for the metric functions $\mathcal{A}(u)$ and $r(u)$ are not available. However, their main qualitative characteristics can be exhibited.
For instance, it is straightforward to check that these functions are smooth in the wormhole section ($-u_{c} < u < u_{c}$). 
Since $A(r)>0$ and $B(r)>0$ for $r_{\textrm{thr}} < r < r_{c}$, it follows from Eq.~\eqref{def-u} that $\textup{d}u / \textup{d} r > 0$ in this interval. Hence, 
$\textup{d} r / \textup{d} u > 0$ for $0 < u < u_{c}$ and 
$\left. \textup{d} r / \textup{d}u \right|_{u=0} = 0$. With the extension beyond the throat, we obtain:
\begin{equation}
\frac{\textup{d} r}{\textup{d} u}\,\,\begin{cases}
<0 & \textrm{, if }-u_{c}<u<0  \\
=0 & \textrm{, if }u=0\\
>0 & \textrm{, if }0<u<u_{c}
\end{cases}
\, .
\label{flare-out-1}
\end{equation}
Result~\eqref{flare-out-1} can be interpreted as the flare-out condition for the wormhole, reflecting the minimum area of the throat \cite{Visser1995}.

For the analysis of $\mathcal{A}(u)$, as $A(r) > 0$ for $r_{\textrm{thr}} < r < r_c$, it follows that $\mathcal{A} > 0$ for $-u_{c}<u<u_{c}$. Considering $\textup{d} \mathcal{A} / \textup{d} u$, we notice that the largest extreme point of $A(r)$ is smaller than $r_{\textrm{thr}}$ \cite{Neves2012}, which implies that 
$\textup{d} A / \textup{d} r < 0$ for $r_{\textrm{thr}} < r < r_{c} \,$. 
Since $\textup{d} \mathcal{A} / \textup{d} u = \left( \textup{d}r/\textup{d}u \right) \, \left( \textup{d} A/ \textup{d}r \right) \,$, 
\begin{equation}
\frac{\textup{d} \mathcal{A}}{\textup{d} u}\,\,\begin{cases}
>0 & \textrm{, if }-u_{c}<u<0\\
=0 & \textrm{, if }u=0\\
<0 & \textrm{, if }0<u<u_{c}
\end{cases}
\, .
\label{comportamento-A}
\end{equation}

\section{Effective potential}
\label{sec:Potential} 

The chart based on the radial coordinate $u$ is not only convenient for extending the geometry beyond the wormhole throat. The $(t,u,\theta,\phi)$ coordinate system also simplifies the analysis of the geodesic dynamics, as we will see in the following.

The equations of motion for geodesics in the geometry of interest can be obtained from the action 
\begin{equation}
I = \int \mathscr{L}\ \textup{d}\lambda  \, ,
   \label{action}
\end{equation}
where the Lagrangian $\mathscr{L}$ is given by
\begin{equation}
- 2 \mathscr{L}
 =  g_{\mu \nu}\frac{\textup{d}x^{\mu}}{\textup{d} \lambda}\frac{\textup{d}x^{\nu}}{\textup{d} \lambda} \, .
\label{eq:lagrangian}
\end{equation}
From Eq.~\eqref{eq:lagrangian}, it follows that $ \mathscr{L}$ is a constant of motion. With a convenient choice of parametrization, it is possible to set $\mathscr{L}$ such that
\begin{equation}
2\mathscr{L} = 
\begin{cases}
0 & \textrm{, for null geodesics} \\
1 & \textrm{, for timelike geodesics}
\end{cases}
\, .
\end{equation}
With the chart $(t,u,\theta,\phi)$, the Lagrangian is written as
\begin{equation}
- 2 \mathscr{L}
=  - \mathcal{A}(u) \left(  \frac{\textup{d}t}{\textup{d}\lambda} \right)^{2}
+ \mathcal{A}(u)^{-1} \left(  \frac{\textup{d}r}{\textup{d}\lambda} \right)^{2} 
+ [r(u)]^{2} \left(  \frac{\textup{d}\theta}{\textup{d}\lambda} \right)^{2} 
+ [r(u)]^{2}\sin^{2} \theta \left(  \frac{\textup{d}\phi}{\textup{d}\lambda} \right)^{2} 
\, .
\label{hamiltonian}
\end{equation}

The conjugated momenta associated with the coordinates $\{ t,u,\theta,\phi \}$ read
\begin{equation}
p_{t} = \mathcal{A}(u) \, \frac{\textup{d}t}{\textup{d}\lambda} \, , \,\,
p_{u} = \mathcal{A}(u)^{-1} \, \frac{\textup{d}r}{\textup{d}\lambda} \, , \,\,
p_{\theta} =  [r(u)]^2 \, \frac{\textup{d}\theta}{\textup{d}\lambda} \, , \,\,
p_{\phi} =  [r(u)]^2\sin^{2} \theta \, \frac{\textup{d}\phi}{\textup{d}\lambda} \, .
\label{conju_moments}
\end{equation}
Hamilton's equations of motion applied to this system imply that there are two additional constants of motion:
\begin{equation}
p_{t} = E = \textrm{constant} \, , \,\, p_{\phi} = L = \textrm{constant} \, .
\label{conserved}
\end{equation}
The integration constants $E$ and $L$ are related to the energy and angular momentum of a given geodesic, respectively. They are a consequence of the staticity and spherical symmetry of the geometry. In addition, it is straightforward to show that a given geodesic is contained in a plane, which can be chosen as $\theta = \pi/2$.

Using the constants of motion and setting $\theta = \pi/2$, the geodesic equations are reduced to
\begin{equation}
\frac{1}{\mathcal{A}(u)} \left(  \frac{\textup{d}u}{\textup{d}\lambda} \right)^{2} 
- \frac{E^2}{\mathcal{A}(u)} + \frac{L^2}{[r(u)]^2}
= - 2 \mathscr{L}
\, ,
\label{eq-geodesics}
\end{equation}
\begin{equation}
\frac{\textup{d}t}{\textup{d}\lambda} = \frac{E}{\mathcal{A}(u)} \, , \,\,
\frac{\textup{d}\phi}{\textup{d}\lambda} = \frac{L}{[r(u)]^{2}} \, .
\label{eq-t-phi}
\end{equation}
Relation~\eqref{eq-geodesics} can be rewritten in the form of a problem in one spatial dimension,
\begin{equation}
\frac{1}{2}  \left(  \frac{\textup{d}u}{\textup{d}\lambda} \right)^{2} + V(u) = \frac{E^{2}}{2} \,\, ,
\label{eq-dinamica-1}
\end{equation}
where the effective potential $V(u)$ is given by
\begin{equation}
V(u) = \mathcal{A}(u) \left\{ 
\frac{L^{2}}{2[r(u)]^{2}} + \mathscr{L}
\right\} \, .
\label{potential}
\end{equation}
In this setup, the dynamics has the state space $\mathcal{M}_1 \equiv \{ u \, | \, -u_{c} < u < u_{c}\}$. From this point onward, we treat the constants $E$ and $L$ as parameters in the dynamical model defined by Eq.~\eqref{eq-dinamica-1}.

An analytical expression for the effective potential $V(u)$ is not available. Nevertheless, its global characteristics can be determined.
For instance, it follows from the definition~\eqref{potential} that $V(u)$ is a smooth function for $u \in \mathcal{M}_1 \,$. 
For radial null geodesics, that is, if $L=0$ and $\mathscr{L} = 0$, $V(u)$ is identically null. 
For any other case ($L \ne 0$ or $\mathscr{L} = 1$), $V(-u_{c}) = V(u_{c}) = 0$ and $V(u) > 0 $ if $-u_{c} < u < u_{c} \, $.

The extrema of $V(u)$ can also be characterized. Let us consider the case where the potential is not identically zero and define the function $\Omega(u)$ as
\begin{equation}
\Omega(u) \equiv
\frac{L^{2}}{2[r(u)]^{2}} + \mathscr{L} \, ,
\label{omega}
\end{equation}
so that the potential can be written as $V(u) = \mathcal{A}(u) \, \Omega(u)$,
and thus,
\begin{equation}
\frac{\textup{d} V}{\textup{d} u} =
 \frac{\textup{d} \mathcal{A}}{\textup{d} u} \, \Omega
+
\mathcal{A} \,\frac{\textup{d} \Omega}{\textup{d} u} \, .
\label{domega}
\end{equation}
Using result~\eqref{comportamento-A}, it is observed that if $0 < u < u_{c} \,$, then $\textup{d} \mathcal{A} / \textup{d} u < 0$ and $\Omega > 0$, and hence $\left( \textup{d} \mathcal{A} / \textup{d} u \right) \Omega < 0 $. 
Also, in this range of $u$, $\mathcal{A} > 0$ and $\textup{d} \Omega / \textup{d} u < 0$, implying that $\mathcal{A} \, \left( \textup{d} \Omega / \textup{d} u \right) < 0$. Combining the previous results and the expression~\eqref{domega}, we obtain that $\textup{d} V / \textup{d} u < 0$.
Similarly, it can be shown that $\textup{d} V / \textup{d} u > 0$ if $-u_{c} < u < 0 \,$. Finally, $\left(\textup{d} \mathcal{A} / \textup{d} u \right) \Omega = 0 $ if $u = 0 \,$. 
Summarizing, if $L \ne 0$ or $\mathscr{L} = 1$,
\begin{equation}
\frac{\textup{d} V}{\textup{d} u}\,\,\begin{cases}
>0 & \textrm{, if }-u_{c}<u<0 \\
=0 & \textrm{, if }u=0\\
<0 & \textrm{, if }0<u<u_{c}
\end{cases}
\, .
\label{comportamento-V}
\end{equation}
And if $L = 0$ and $\mathscr{L} = 0$, then $\textup{d} V / \textup{d} u = 0\,$ for $-u_{c} < u < u_{c} \,$.

Previous considerations furnish an overall picture of the effective potential associated with the geodesic dynamics in the de Sitter wormhole. For radial null geodesics, the potential is identically null. For any other case, the potential is zero at the cosmological horizons and positive elsewhere, with a single maximum at the throat.

Complementing the qualitative analysis, an analytical expression for $V(u)$ can be obtained near the throat. This region is important for geodesic dynamics.
The function $r(u)$ is smooth around $u=0$, with
\begin{equation}
\left. \frac{\textup{d} r}{\textup{d} u}\right|_{u = 0} = 0 \, , \,\,
\left. \frac{\textup{d}^2 r}{\textup{d} u^2}\right|_{u = 0} = 2K > 0
\, ,
\label{flare-out}
\end{equation}
where the (positive) constant $2K$ is expressed in terms of the wormhole parameters as 
\begin{gather}
2K  = \frac{C - 1}{2} \left. \frac{dP(r)}{dr} \right|_{r_{\textrm{thr}}}  \nonumber \\
 =  \frac{(1 - C) P(r_{\textrm{thr}}) }{2}  \left[
\frac{c_0}{r_{\textrm{thr}} - r_{0}}
+ \frac{c_{0-}}{r_{\textrm{thr}} - r_{0-}}
+ \frac{c_{0--}}{r_{\textrm{thr}} - r_{0--}}
+ \frac{c_{0n}}{r_{\textrm{thr}} - r_{0n}}
\right] \, .
\label{def-K}
\end{gather}
Using Eq.~\eqref{flare-out}, the metric functions $r(u)$ and $\mathcal{A}(u)$ are written as
\begin{eqnarray}
r(u) & = & r_{\textrm{thr}} + K \, u^{2} +   O \left( u^{3} \right)\,  ,
\label{approx-r} \\
\mathcal{A} (u) & = & \mathcal{A}_{0} + \mathcal{A}_{2} \, u^{2} +  O \left( u^{3} \right) \, ,
\label{approx-A}
\end{eqnarray}
with the Taylor coefficients,
\begin{equation}
\mathcal{A}_{0} \equiv  A(r_{\textrm{thr}}) \, ,
\,\,
\mathcal{A}_{2} \equiv
K \left. \frac{dA}{dr}  \right|_{r = r_{\textrm{thr}}}  \, .
\label{def-A2}
\end{equation}
Substituting results~\eqref{approx-r} and \eqref{approx-A} into Eq.~\eqref{potential}, an expression for the effective potential is obtained,
\begin{equation}
V(u) = V_{0} + V_{2} \, u^{2} + O \left( u^{3} \right)
\, ,
\label{approx-V}
\end{equation}
with
\begin{equation}
V_{0} \equiv \mathcal{A}_{0} \left( 
\frac{L^{2}}{2 r_{\textrm{thr}}^{2}} + \mathscr{L}
\right) \, , \,\,
V_{2} \equiv
\mathcal{A}_{2} \left( 
\frac{L^{2}}{2 r_{\textrm{thr}}^{2}} + \mathscr{L}
\right)
- \frac{\mathcal{A}_{0} L^{2} K}{r_{\textrm{thr}}^{3}}
\, . 
\label{def-V0V2}
\end{equation}
The constant $V_{0}$ represents a background contribution to the potential, while the term $V_{2}$ measures a perturbation in this background.

As a consistency check, we will verify that $V_{2}$ is non-null and negative if $L \ne 0$ or $\mathscr{L} = 1$. From Eq.~\eqref{domega},
\begin{equation}
2V_{2} = 
\Omega(0) \, \left. \frac{\textup{d}^{2} \mathcal{A}}{\textup{d} u^{2}}  \right|_{u=0}
+
\mathcal{A}(0) \, \left. \frac{\textup{d}^{2} \Omega}{\textup{d} u^{2}}  \right|_{u=0}
+ 
\left. \frac{\textup{d}\mathcal{A}}{\textup{d} u}  \right|_{u=0}
\,
\left. \frac{\textup{d}\Omega}{\textup{d} u}  \right|_{u=0}
\, .
\label{ddomega}
\end{equation}
Assuming $L \ne 0$ or $\mathscr{L} = 1$, 
$\mathcal{A}(0)>0$, 
$\Omega(0)>0$, 
$\textup{d} \mathcal{A} / \textup{d} u |_{u=0} = \textup{d} \Omega / \textup{d} u |_{u=0} = 0$, 
$\textup{d}^{2} \mathcal{A} / \textup{d} u^{2} |_{u=0} < 0$ 
and 
\linebreak
$\textup{d}^{2} \Omega / \textup{d} u^{2} |_{u=0} < 0$.
It follows from Eq.~\eqref{ddomega} that
\begin{equation}
V_{2} < 0 \, .
\label{V2negative}
\end{equation}

\section{\label{sec:Dynamics} Dynamical analysis}
\label{sec:dynamics}

\subsection{Effective dynamical system}
\label{subsec:dynamical-system}

Spacetime symmetries simplified the system of interest into an effective one-dimensional problem characterized by Eq.~\eqref{eq-geodesics}. The introduction of the quasilocal coordinate $u$ further simplified the analysis, allowing the definition of the effective potential~\eqref{potential}. 
In fact, from the solution of Eq.~\eqref{eq-dinamica-1}, the functions $t(\lambda)$ and $\phi(\lambda)$ in Eq.~\eqref{eq-t-phi} are readily integrable. 
Nevertheless, the quadratic kinetic term in Eq.~\eqref{eq-dinamica-1} leads to some complications in treatment. To overcome this problem, the strategy proposed in the present work is to transform the one-dimensional equation~\eqref{eq-dinamica-1} into a two-dimensional dynamical system with a constraint.

Differentiating the relation~\eqref{eq-dinamica-1} with respect to $\lambda$, an autonomous differential equation of second order is obtained:
\begin{equation}
\frac{\textup{d}^{2} u}{\textup{d}\lambda^{2}} = 
- \frac{\textup{d}V}{\textup{d}u} \, .
\label{geodesic_equation}
\end{equation}
Using Eq.~\eqref{geodesic_equation}, we rewrite the dynamical system defined by Eq.~\eqref{eq-dinamica-1} with two first-order differential equations
\begin{eqnarray}
\frac{\textup{d} u}{\textup{d}\lambda} & = &  w \, , 
\label{system_eq-1}\\
\frac{\textup{d} w}{\textup{d}\lambda} & = & - \frac{\textup{d}V}{\textup{d}u} \, ,
\label{system_eq-2}
\end{eqnarray}
equipped with the constraint
\begin{equation}
\frac{1}{2} w^{2} + V(u) = \frac{E^{2}}{2} \, .
\label{constraint}
\end{equation}
The state space of the two-dimensional dynamical system given by Eqs.~\eqref{system_eq-1}--\eqref{system_eq-2} is denoted by $\mathcal{M}_{2}$. The dynamical system of interest, taking into account the constraint~\eqref{constraint}, has the state space $\tilde{\mathcal{M}}_{1} = \{ (u,w) \, | \, \nicefrac{w^{2}}{2} + V(u) = \nicefrac{E^{2}}{2} \}$. Note that, although $\tilde{\mathcal{M}}_{1} \subset \mathcal{M}_{2} \,$, the state spaces $\tilde{\mathcal{M}}_{1}$ and $\mathcal{M}_{1}$ are homeomorphic.

Another interpretation of the constraint~\eqref{constraint} is that it describes the possible orbits of the dynamical system in the two-dimensional state space $\mathcal{M}_{2}$. That is, different values of the energy $E$ and angular momentum $L$ correspond to different copies of $\tilde{\mathcal{M}}_{1}$, each representing a geodesic in the wormhole spacetime.

The fixed points of the dynamical system can be determined. Let us denote a fixed point by $(u_{\star},w_{\star})$. Taking into account Eqs.~\eqref{system_eq-1}--\eqref{system_eq-2}, we have
\begin{eqnarray}
w_{\star} & = & 0 \, ,
\label{fixed-point-cond1}\\
\left. \frac{\textup{d}V}{\textup{d}u} \right|_{u_{\star}} & = & 0 \, .
\label{fixed-point-cond2}
\end{eqnarray}
Additionally, a fixed point must satisfy the constraint~\eqref{constraint}. Together with Eq.~\eqref{fixed-point-cond1}, Eq.~\eqref{constraint} implies
\begin{equation}
V(u_{\star}) = \frac{E^{2}}{2} \, .
\label{fixed-point-cond3}
\end{equation}

Considering the cases where $L \ne 0$ or $\mathscr{L} = 1$,
Eq.~\eqref{fixed-point-cond2} shows that a fixed point is an extreme point of the effective potential. However, in Sec.~\ref{sec:Potential} we have determined that there is only one extreme of $V(u)$, namely the point $u=0$ (a maximum). But condition~\eqref{fixed-point-cond3} also has to be satisfied. Taking into account that $V(0) = V_{0}$, 
\begin{equation}
\left( u_{\star}, w_{\star} \right) = (0,0) \,\, \textrm{with } E = \pm \sqrt{2V_{0}} \,\,\,  (L \ne 0 \textrm{  or  } \mathscr{L} = 1) \, ,
\label{fixed-point}
\end{equation}
with the constant $V_{0}$ given by Eq.~\eqref{def-V0V2}. 

On the other hand, if $L = 0$ and $\mathscr{L} = 0$, i.e. radial null geodesics, condition~\eqref{fixed-point-cond3} implies that $E=0$. Therefore, there are only simple solutions associated with this case and no isolated fixed point. This scenario will be discussed further in Sec.~\ref{sec:radial-null-geodesics}.

\subsection{Stability tests}
\label{sec:stability}

The goal of this section is to investigate the stability of the photon sphere, which is interpreted here as a fixed point of the constructed dynamical system.
The literature on dynamical systems presents several notions of stability for characterizing fixed points. However, these different notions often do not agree in many systems \cite{boehmer2012,sabau2005}, which highlights the need for systems that show agreement between more than one stability notion. In this subsection, we will introduce the Lyapunov and Jacobi stability notions. The strategy used is to work in the two-dimensional state space $\mathcal{M}_{2}$, complemented by the fixed-point condition~\eqref{fixed-point}.

The effective dynamical system given by Eqs.~\eqref{system_eq-1}--\eqref{constraint} can be locally characterized by its linear stability. This method is based on the linearization of the dynamics and the study of the eigenvalues of the Jacobian matrix associated with the fixed point $(u_{\star},w_{\star}) = (0,0)$. Let $\{ \nu_{i} \}$ be the eigenvalues associated with this Jacobian. This fixed point is said to be Lyapunov unstable if (at least) one eigenvalue $\nu_{\text{i*}}$ has a positive real part [$\textrm{Re}(\nu_{\text{i*}}) > 0$], and Lyapunov stable otherwise. 

Considering the dynamics near the wormhole throat, we have
\begin{equation}
    u = u_\star + \delta u = \delta u \,\, , \qquad  w = w_\star + \delta w = \delta w \,\, ,
    \label{eq:perturbation}
\end{equation}
and
\begin{equation}
   \frac{\textup{d}}{\textup{d}\lambda}\begin{pmatrix}
       \delta u \\
       \delta w
   \end{pmatrix} = 
    \mathbb{J} \, 
    \begin{pmatrix}
        \delta u \\
        \delta w
    \end{pmatrix}\, ,
%\label{eq:jacobian}
\end{equation}
with the Jacobian matrix $\mathbb{J}$ given by
\begin{equation}
    \mathbb{J} = 
    \begin{pmatrix}
        0 & 1 \\
        - \frac{\textup{d}^{2} V}{\textup{d}u^{2}} & 0
    \end{pmatrix}_{
    \substack{
    (u , w ) = (0,0)
    \\
    E^{2} = 2 V_{0}
    }
    } = 
    \begin{pmatrix}
        0 & 1 \\
        -2V_{2} & 0 
    \end{pmatrix} \, .
\label{eq:jacobian}
\end{equation}
The associated (real) eigenvalues $\nu_{+}$ and $\nu_{-}$ are
\begin{equation}
    \nu_{\pm} = \pm \sqrt{-2V_{2}} \, .
    \label{eq:eigenvalue}
\end{equation}
Since $V_{2}<0$, as seen in Eq.~\eqref{V2negative}, the dynamic is hyperbolic, and therefore the linear analysis is appropriate. Because $\nu_{+}>0\,$, the fixed point is Lyapunov unstable.

Another characterization of stability is the Jacobi criterion, which is associated with deviations of nearby orbits around a fixed point in the state space.
To present this notion of stability, we will introduce the relevant quantities and restrict the formalism to the present wormhole geometry that we study in this work. Initially, consider the following second-order ordinary differential equation in the standard form \cite{boehmer2012,sabau2005,hossein2012}:
\begin{equation}
    \frac{\dder^2 u}{\dder \lambda^2} = -2G(u,w) \, ,
    \label{eq:soode}
\end{equation}
where $w = \dder u \, / \dder \lambda$ and with an associated fixed point in $u = u_{\star} = 0$. 
The concept of a fully covariant differentiation along the trajectories of the second-order differential equation~\eqref{eq:soode} is captured in the Kosambi-Cartan-Chern (KCC) framework. This approach takes into account corrections due to the nonlinearity of the system. Let $\xi$ denote the first-order variation of $u$. The KCC-covariant derivative $\Dder \xi /\dder \lambda$ of the variation field $\xi$ incorporates both the affine connection of the manifold and the nonlinear connection induced by the dynamical system. Explicitly, this derivative has the form \cite{boehmer2012, sabau2005},
\begin{equation}
    \frac{\Dder \xi}{\dder \lambda} = \frac{\dder \xi}{\dder \lambda} + \frac{\partial G}{\partial w} \xi \, ,
\end{equation}
where $\partial G/\partial w$ is the nonlinear connection. Varying the trajectories in Eq.~\eqref{eq:soode} into nearby ones, one gets the Jacobi equation, 
\begin{equation}
    \frac{\Dder^2 \xi}{\dder \lambda^2} = P(u,w) \xi \, ,
\end{equation}
where $P(u,w)$ is the deviation curvature scalar, defined by
\begin{equation}
    P(u,w) = -2\frac{\partial G}{\partial u} -2 G \tilde{G} +w\frac{\partial N}{\partial u} + N^2 \, ,
\end{equation}
with the quantities
\begin{equation}
    N = \frac{\partial G}{\partial w} \, , \quad \tilde{G} = \frac{\partial N}{\partial w} \, ,
\end{equation}
being the nonlinear and Berwald connections, respectively \cite{boehmer2012}.
Jacobi stability depends on $P(u,w)$ evaluated at the fixed point $(u_{\star} , w_{\star} ) = (0,0)$. Specifically, if $P(u_{\star}, w_{\star}) < 0$ the trajectories in Eq.~\eqref{eq:soode} are stable, if $P(u_{\star}, w_{\star}) > 0$ they are unstable. The marginal case $P(u_{\star}, w_{\star}) = 0$  must be studied case by case.

From the geodesic equation~\eqref{geodesic_equation}, one can identify,
\begin{equation}
    G(u,w) = \frac{1}{2}V'(u) \, ,
\end{equation}
with associated connections $N = 0$  and $\tilde{G} = 0$. These results furnish the following deviation curvature scalar,
\begin{equation}
      \left.  P(u_\star, w_{\star}) \right|_{(u_\star, w_\star) = (0,0)} = -V''(0) = -2V_2 \, .
\end{equation}
As seen in Eq.~\eqref{V2negative}, $V_{2}$ is negative and hence $P(0, 0)>0$. Therefore, the trajectories are Jacobi unstable. This result shows that both notions of stability, Jacobi and Lyapunov, are in agreement. This is a relevant feature, observed in other brane geometries and cosmological solutions \cite{Klen2020, boehmer2012}.

\subsection{Null geodesics and the photon sphere}
\label{sec:null-geodesics}

After the general discussion presented in previous sections, we consider specific dynamical properties of null geodesics ($\mathscr{L} = 0$) in the de Sitter wormhole, initially focusing on nonradial geodesics ($L \ne 0$).

The fixed point in the dynamical system is associated with geodesics whose orbits are closed, with a constant radius. The surface generated by these orbits is the so-called photon sphere \cite{khoo2016, claudel2001}.
Result~\eqref{fixed-point} indicates that the wormhole throat ($u=0$ or $r=r_{\textrm{thr}}$) is the unique photon sphere in space-time.

Let us examine the photon-sphere condition. With $L\ne 0$ and $\mathscr{L} = 0$, the constraint~\eqref{fixed-point-cond3} furnishes 
\begin{equation}
E^{2} = 2 V_{0}
\Longrightarrow 
L^{2} =
\frac{r_{\textrm{thr}}^{2} E^{2}}{\mathcal{A}_{0}}
 \, .
\label{eq:system_imp}
\end{equation}
Expression~\eqref{eq:system_imp} can be interpreted as the condition for a light ray to be located on the wormhole throat (the photon sphere). Introducing the impact parameter ($D$), defined as
\begin{equation}
D \equiv \frac{L}{E} \, ,
\label{eq:impact}
\end{equation}
the photon-sphere condition~\eqref{eq:system_imp} is written as
\begin{equation}
D = \pm D_{\textrm{crit}} \, ,
\end{equation}
with the critical impact parameter $D_{\textrm{crit}}$ given by
\begin{equation}
D_{\textrm{crit}} \equiv \frac{r_{\textrm{thr}}}{\sqrt{\mathcal{A}_0}} \, .
\label{eq:Dcrit}
\end{equation}

\begin{figure}[ht]
\includegraphics[width=.4\textwidth]{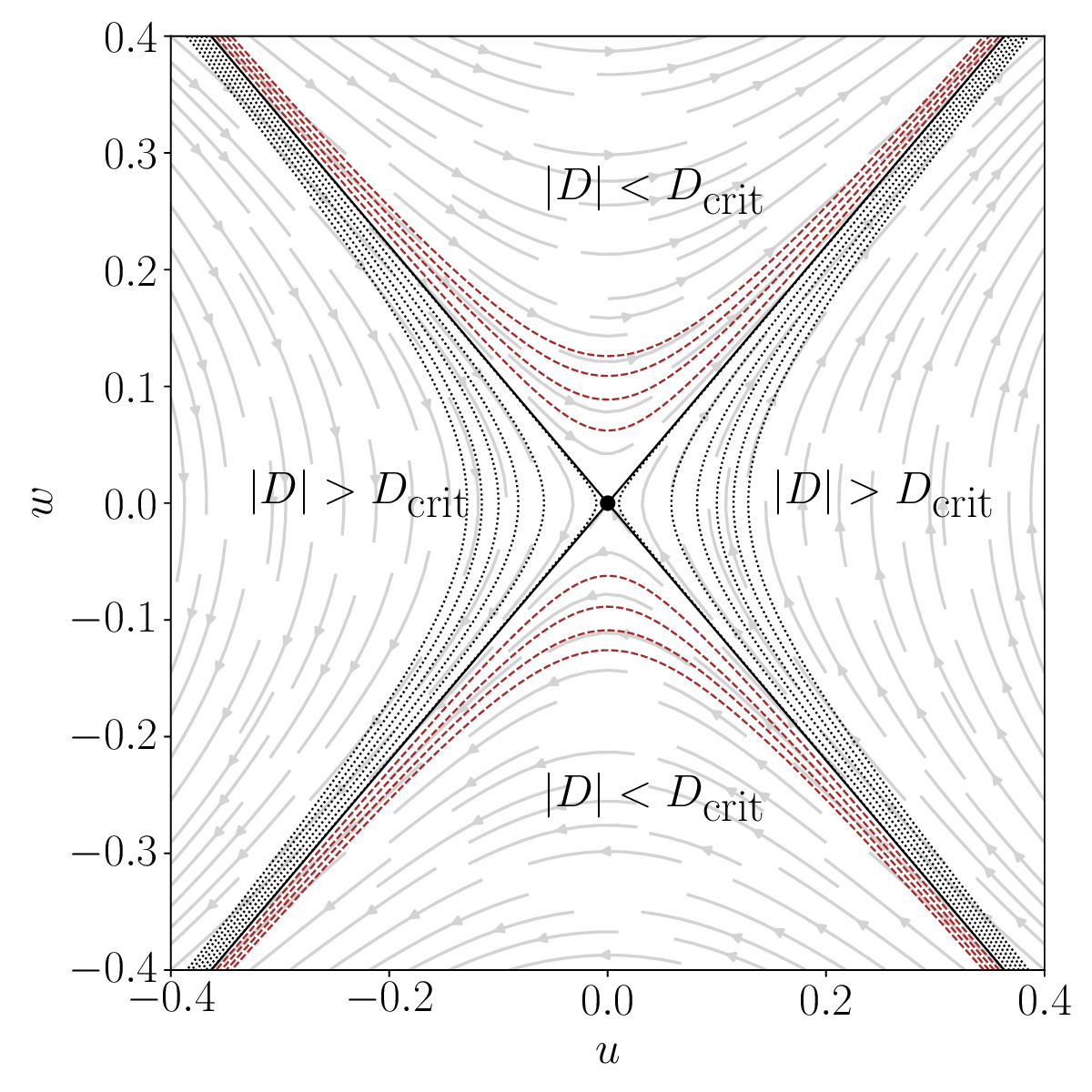}
\caption{Typical orbits in the two-dimensional phase portrait $\mathcal{M}_{2}$ of the null-geodesic case, for different values of the impact parameter. Dashed (dotted) lines represent trajectories that cross (do not cross) the throat. Dashed and dotted lines can also be interpreted as copies of $\tilde{\mathcal{M}_{1}}$. In this graph, $M=1.0$,  $q = 0.5$, $\Lambda = 0.1$, $C = 0.5$, and several values of $D$.}
\label{ap:fig:phase_space}
\end{figure}

If the magnitude of the null geodesic's impact parameter is less than the critical value ($|D| < D_{\textrm{crit}}$), the geodesic crosses the throat, going from one patch of the wormhole to the other (dotted lines in Fig.~\ref{ap:fig:phase_space}). Otherwise, if $|D| > D_{\textrm{crit}}$, there is a turning point in the trajectory, and the geodesic bounces back in the same patch (dashed lines in Fig.~\ref{ap:fig:phase_space}). 
The critical case $ | D | =  D_{\textrm{crit}} $ corresponds to photons asymptotically approaching the throat in infinite affine time, forming an unstable photon sphere.

\begin{figure}
\includegraphics[width=0.4\linewidth]{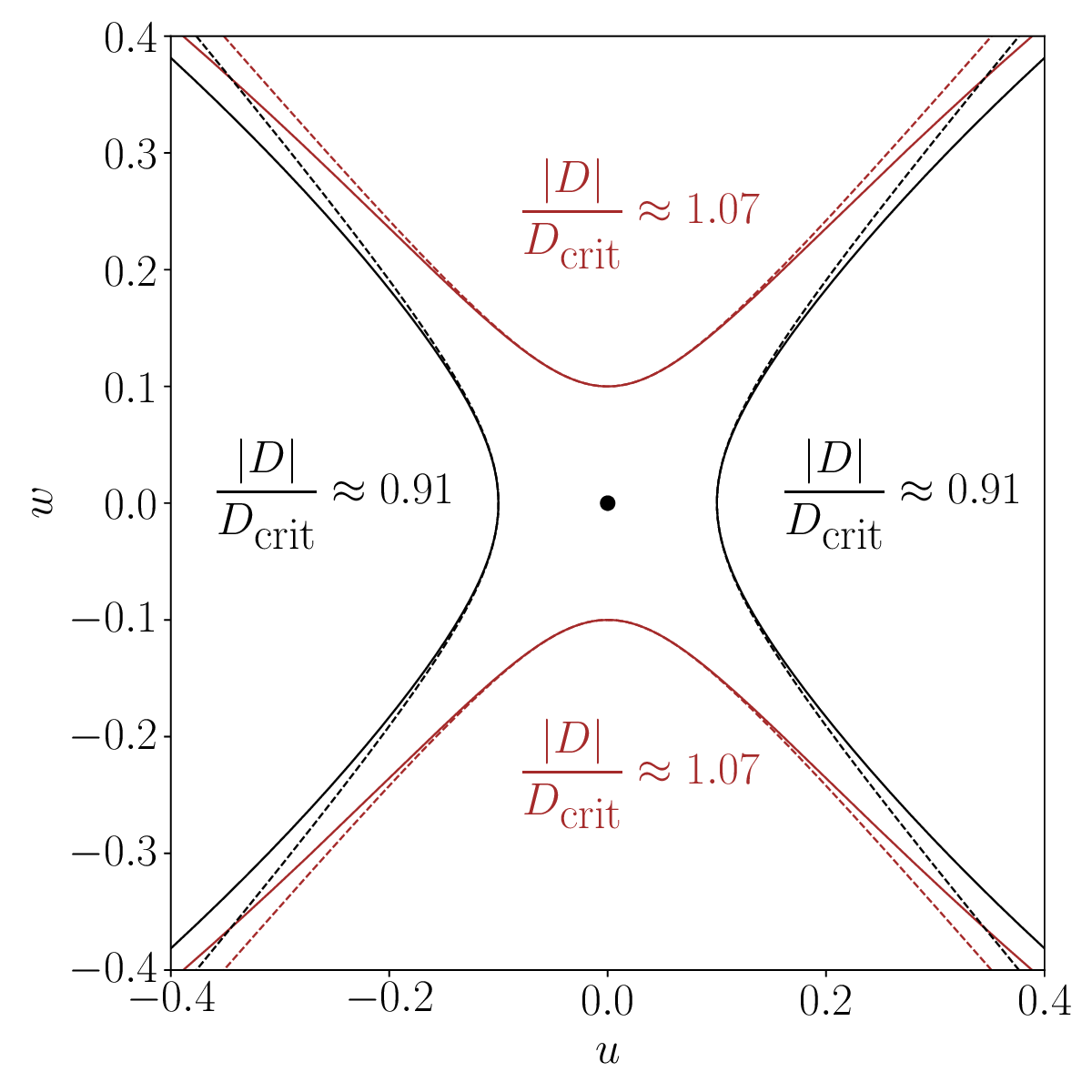}
\caption{Comparison between the approximate expression (solid lines) and the numerical results (dashed lines) for orbits close to the fixed point of the null-geodesic dynamics. In this graph, $M=1.0$,  $q = 0.5$, $\Lambda = 0.1$, $C = 0.5$, and several values of $D$.}
    \label{fig:desv_luz}
\end{figure}

An analytical solution for the null geodesics is available around the wormhole throat, that is, near the origin of the two-dimensional state space $\mathcal{M}_{2}$ with $D \approx D_{\text{crit}}$. In this regime, the solution, with respect to the initial position $\left( u(0) , w(0) \right)$, can be written as
\begin{eqnarray}
u(\lambda) & = & u(0) \cosh \left(\sqrt{2|V_2|} \lambda\right)+\frac{w(0) \sinh \left(\sqrt{2|V_2|}\lambda\right)}{\sqrt{2|V_2|}} \, , 
\label{eq:sol1_hiper} \\
w(\lambda) & = & u(0)\sqrt{2|V_2|}\sinh \left(\sqrt{2|V_2|}\lambda \right) + w(0)
   \cosh \left(\sqrt{2|V_2|} \lambda \right)  \, ,
\label{eq:sol2_hiper}
\end{eqnarray}
with $V_{2}$ defined in Eq.~\eqref{def-V0V2}. 
Equations~\eqref{eq:sol1_hiper}--\eqref{eq:sol2_hiper} describe hyperbolic trajectories in $\mathcal{M}_{2} \,$, which can be written as
\begin{equation}
    w^2 
    - E^{2} \left(
    \frac{2  K}{r_{\textrm{thr}}} \, 
    - \frac{\mathcal{A}_{2}}{\mathcal{A}_{0}} 
   \right) \, u^{2}
    =  
    2 E^{2} \left( 1 -  \frac{|D|}{D_{\text{crit}}} \right ) 
    \, .
    \label{constraint2}
    \end{equation}
The hyperbole in Eq.~\eqref{constraint2} approximates the orbits shown in Fig.~\ref{ap:fig:phase_space}. This expression can also be interpreted as the constraint in Eq.~\eqref{constraint} near the fixed point of the dynamics.
There is a noticeable improvement in the quality of the approximation as the trajectory approaches the wormhole's throat, as seen in  Fig.~\ref{fig:desv_luz}.

By reparametrizing the radial differential equation in terms of the angular variable $\phi$, trajectories~\eqref{eq:sol1_hiper}--\eqref{eq:sol2_hiper} are identified as null geodesics spiraling either into or out of the wormhole. In the critical case, when the parameter $D$ is equal to its critical value $D_{\textrm{crit}}$, the null geodesic corresponds to a closed circular orbit at the wormhole throat. It corroborates the fact that the throat is an unstable photon sphere, where light rays are trapped for some time but eventually escape.

\subsection{Radial null geodesics and bifurcation}
\label{sec:radial-null-geodesics}

Radial null trajectories ($L=0$ and $\mathscr{L} = 0$) are presently treated. From Eq.~\eqref{eq-t-phi}, we obtain
\begin{equation}
\frac{\textup{d}\phi}{\textup{d}\lambda} = 0 \Longrightarrow 
\phi(\lambda) = \textrm{constant}
\, .
\end{equation}
For this class of null geodesics, the effective potential is identically null, and the fixed-point condition~\eqref{fixed-point} is not satisfied. Nevertheless, the dynamical system can be solved exactly, with
\begin{equation}
u(\lambda) = \pm E \lambda + u_{0} \, .
\end{equation}
The integration constant $u_{0}$ can be interpreted as an initial position.

We observe that the topology of the phase portrait changes drastically from the case of radial null geodesics to the case of nonradial geodesics. When dealing with radial geodesics, one sets $L = 0$, which makes the effective potential in Eq.~\eqref{approx-V} identically zero. 
In this scenario, a continuous line of fixed points $\{ (u_{\star}, 0), - u_{c} < u_{\star} < u_{c} \}$ appears in the system.
On the other hand, when considering nonradial null geodesics, one sets $L\neq 0$, which makes the effective potential quadratic and gives an isolated fixed point in the dynamics. This behavior shows a change in the topology of the phase portrait of the system, indicating the existence of a bifurcation when $L= 0$.

The Jacobian matrix $\mathbb{J}$ in Eq.~\eqref{eq:jacobian} for null geodesics at the critical value $L = 0$ is
\begin{equation}
\mathbb{J} = 
\begin{pmatrix}
0 & 1 \\
0 & 0
\end{pmatrix} 
\Longrightarrow 
\det \mathbb{J} = 0
\, .
\label{eq:deg_jac}
\end{equation}
The eigenvalues exhibit degeneracy with double multiplicity, and the matrix $\mathbb{J}$ has only a single eigenvector. That is, when $L=0$ and $\mathscr{L} = 0$, the eigenspace corresponding to $\mathbb{J}$ becomes one dimensional. This behavior is indicative of a codimension-two Bogdanov-Takens bifurcation, as discussed in \cite{Klen2020, kuznetsov2005, Azim2020}. Such a bifurcation is characterized by the coexistence of a double-zero eigenvalue and a one-dimensional eigenspace, which plays a critical role in the dynamical behavior of the system.

\subsection{Null geodesics and wormhole shadow}
\label{sec:shadow}

Presently, we consider the optical appearance of the wormhole to a static observer. More specifically, we will study the so-called ``shadow'' of the wormhole \cite{Perlik2018,Roy2020} as seen by an observer in a fixed position, $u = u_{\odot}$, near the wormhole throat ($u_{\odot} \approx 0$). 
We also assume that the observer's angular coordinates are $\theta = \theta_{\odot} \equiv \pi/2$ and $\phi = \phi_{\odot} \equiv 0$.
\begin{figure}[ht]
\includegraphics[width = 0.35 \textwidth]{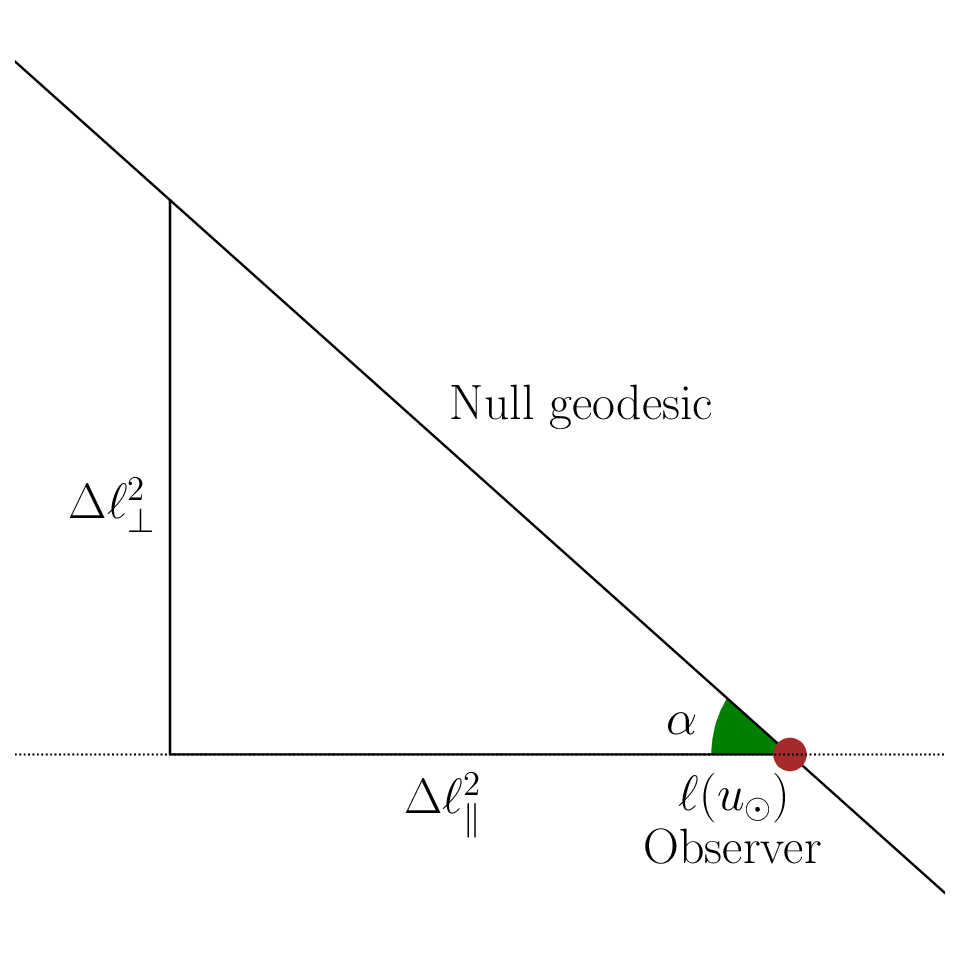}
\caption{Light ray received by an observer and relevant quantities for the analysis of the wormhole shadow.}
\label{fig:gravitationalLens}
\end{figure}

Let us consider, without loss of generality, that a light ray reaches the observer following a trajectory in the plane $\theta = \pi/2$. From Eq.~\eqref{metric_u}, the line element in this 3-surface can be written as
\begin{equation}
\dder s^2_{3} = -\mathcal{A}(u) \dder t^2 + \dder \ell_{\parallel}^{2} + \dder \ell_{\perp}^{2} \, ,
\end{equation}
where
\begin{equation}
\dder \ell_{\parallel} = \frac{ \dder u}{[\mathcal{A}(u)]^{1/2}} \, , \,\,
\dder \ell_{\perp} = r(u) \dder \phi \,.
\label{ell_paralelo_perpendicular}
\end{equation}
The scalar $\ell_{\parallel}$ is a proper coordinate of the observer whose axis is parallel to the radial axis. Conversely, $\ell_{\perp}$ is a proper coordinate perpendicular to the radial coordinate, in the plane $\theta = \pi/2$.

Denoting by $\alpha$ the angle of the light ray with the radial direction, we have
\begin{equation}
\tan \alpha = \frac{\Delta \ell_{\perp}}{\Delta \ell_{\parallel}} \, ,
\label{def-alpha}
\end{equation} 
where $\Delta \ell_{\perp}$ and $\Delta \ell_{\parallel}$ are the proper lengths of the triangle in Fig.~\ref{fig:gravitationalLens}, defined by the trajectory of the light ray near the observer at $u=u_{\odot}$.

Using Eqs.~\eqref{ell_paralelo_perpendicular} considering the limit near the observer,
\begin{equation}
\frac{\Delta \ell_{\perp}}{\Delta \ell_{\parallel}} \longrightarrow
\left. 
\frac{r(u)}{[\mathcal{A}(u)]^{1/2}} \frac{\dder \phi}{\dder u}
\right|_{u=u_{\odot}} \, ,
\end{equation}
hence Eq.~\eqref{def-alpha}, combined with the approximations in Eqs.~\eqref{approx-r} and \eqref{approx-A}, gives
\begin{equation}
    \tan^2(\alpha) = \left. \frac{(r_{\textrm{thr}}+Ku^2)^2}{(\mathcal{A}_0 +\mathcal{A}_2 u^2)^{-1}}\left(\frac{\textup{d}\phi}{\textup{d}u} \right)^2 \right|_{u = u_\odot} \, .
\end{equation}

The derivative $\textup{d}\phi/\textup{d}u$ at $u=u_{\odot}$ can be found from the geodesic equations~\eqref{eq-geodesics}-\eqref{eq-t-phi},
\begin{equation}
    \left( \frac{\textup{d}\phi}{\textup{d}u}\right)^2 = \frac{D^2}{r(u)^2 [r(u)^2-\mathcal{A}(u) D^2]} \, ,
    \label{eq:dfidu}
\end{equation}
where $D$ is the impact parameter defined in Eq.~\eqref{eq:impact}, and consequently,
\begin{equation}
    \tan^2 (\alpha) =  \frac{D^{2} \left(\mathcal{A}_{0} + \mathcal{A}_{2} u_\odot^2\right)}{\left(K u_\odot^{2} + r_{\textrm{thr}}\right)^{2} - D^{2} \left(\mathcal{A}_0+\mathcal{A}_2 u_\odot^2\right)}  \, .
\end{equation}
Using the identity $\sin^2(\alpha) = \tan^2(\alpha)[ \tan^2(\alpha) - 1]^{-1}$ and the result~\eqref{eq:Dcrit}, the following expression for the angle $\alpha$ is obtained:
\begin{equation}
    \sin^2(\alpha) = \frac{r_\textrm{thr}^2 \left(\mathcal{A}_0+\mathcal{A}_2 u_\odot^2\right)}{2 r_{\textrm{thr}} u_\odot^2 (\mathcal{A}_2 r_{\textrm{thr}}-\mathcal{A}_0 K)-\mathcal{A}_0 K^2 u_\odot^4 + \mathcal{A}_0
   r_\textrm{thr}^2} \, .
\label{eq:StaticShadow}
\end{equation}
The boundary of the shadows is traced by light rays whose angle $\alpha$ at the observer is given by Eq.~\eqref{eq:StaticShadow}.

As a consistency check, we verify from Eq.~\eqref{eq:StaticShadow} that
\begin{equation}
    \lim_{u_\odot \to 0} \sin^2(\alpha) = 1 \, ,
\end{equation}
as it should be.
This result implies that $\alpha = \pi/2$ for an observer located at the throat $u=0$ (which is a photon sphere). 
At this position, half of the sky is illuminated by the stars on one side of the wormhole, while the other half is illuminated by the stars on the other side of the wormhole. 
We emphasize that our shadow analysis applies only to observers close to the wormhole throat, since the derived expressions are valid only in this near-throat regime. This differs from standard shadow analyses in the literature, which typically consider distant observers, particularly for symmetric wormholes with a single photon sphere \cite{Ohgami2015,Bouhmadi2021}. Our near-field treatment reveals distinctive observational signatures that complement the well-studied far-field case.

\subsection{Timelike geodesics}
\label{sec:generalDynamics}

\begin{figure}[ht]
\includegraphics[width =.4\textwidth]{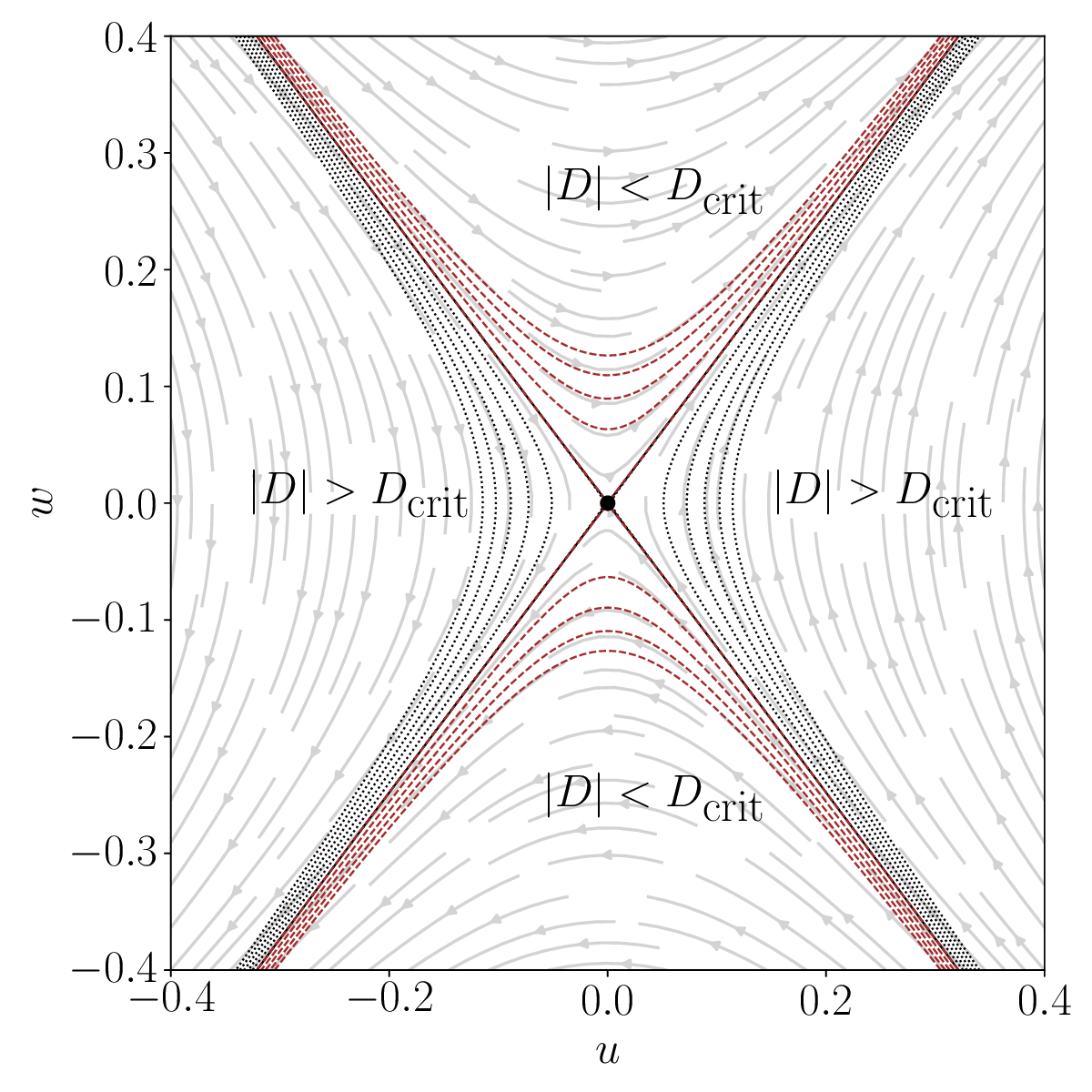}
\caption{Two-dimensional phase portrait $\mathcal{M}_{2}$ for the timelike dynamical system with $M=1.0$,  $q = 0.5$, $\Lambda = 0.1$, $C = 0.5$, and several values of $D$.}
\label{fig:spacetimelike}
\end{figure}

Let us consider the dynamics of timelike geodesics ($\mathscr{L}=1$) in the background of the de Sitter wormhole. As noted in Sec.~\ref{sec:Potential}, the effective potential~\eqref{potential} is symmetric and has a single maximum point around $u=0$. Therefore, general results derived in Secs.~\ref{subsec:dynamical-system}--\ref{sec:stability} indicate that null and timelike geodesic dynamics are qualitatively similar. In particular, the only fixed point of the dynamical system is $(u_{\star}=0,w_{\star}=0)$, that is, at the wormhole throat. It is an unstable fixed point according to both Lyapunov and Jacobi criteria. A typical phase diagram for the dynamics of timelike geodesics is shown in Fig.~\ref{fig:spacetimelike}. 

With the points discussed, one can sketch the motion of a massive particle traveling in the direction of the wormhole. 
If the particle has enough energy ($E > V_{0}$), it crosses the wormhole throat towards another patch in the direction of the cosmological horizon. 
For the critical energy value $E = V_{0}$, the particle is temporarily trapped in a closed circular orbit around the throat. 
And for a low-energy particle ($E < V_{0}$), it bounces back without reaching a bounded orbit.
 
\begin{figure}
\includegraphics[width=0.4\linewidth]{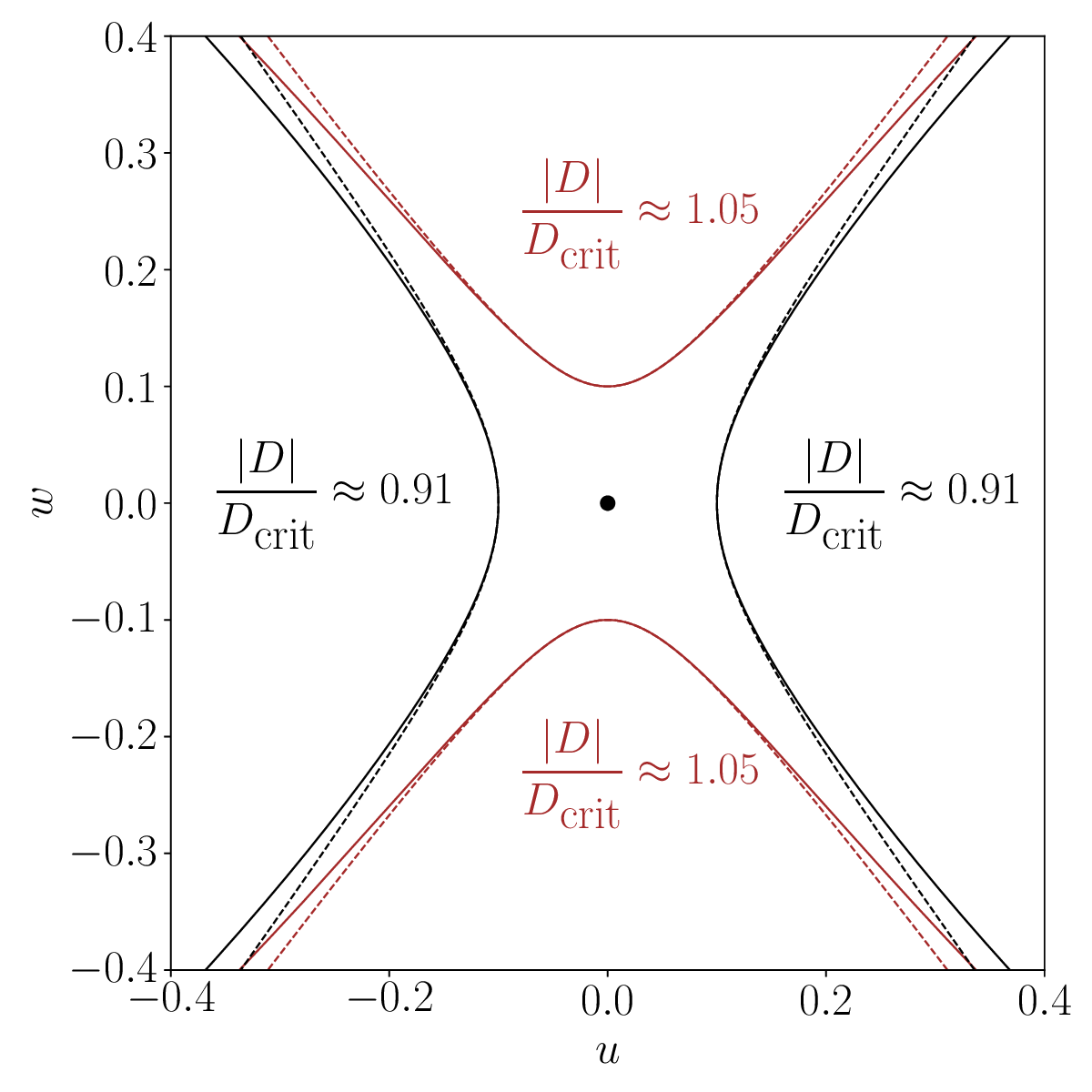}
\caption{Comparison between the approximate expression (solid lines) and the numerical results (dashed lines) for orbits close to the fixed point of the timelike-geodesic dynamics. In this graph, $M=1.0$,  $q = 0.5$, $\Lambda = 0.1$, $C = 0.5$, and several values of $D$.}
    \label{fig:desv_tempo}
\end{figure}

The general behavior of massive particles in the wormhole spacetime is captured by their trajectories near the throat. Using results~\eqref{eq:jacobian}--\eqref{eq:eigenvalue}, analytical expressions can be derived. For $\mathscr{L}=1$, this local solution can be expressed as
\begin{equation}
    \begin{pmatrix}
    u(\lambda) \\
    w(\lambda)
    \end{pmatrix} = 
   C_{+} \begin{pmatrix}
        \sqrt{2 \left| V_{2} \right|} \\
        1
    \end{pmatrix} e^{ \sqrt{2 \left| V_{2} \right|} \, \lambda} 
  + C_{-} \begin{pmatrix}
        - \sqrt{2 \left| V_{2} \right|} \\
        1
    \end{pmatrix} e^{- \sqrt{2 \left| V_{2} \right|} \, \lambda} \, ,
    \label{eq:linearizedSol}
\end{equation}
with the constant $V_{2}$ in Eq.~\eqref{def-V0V2} depending on the wormhole parameters, and the constants $C_{\pm}$ that can be determined from the initial condition.
The solution~\eqref{eq:linearizedSol} is associated to the hyperbolic orbits in state space $\mathcal{M}_{2}$ given by
\begin{equation}
    w^2 
    - E^{2} \left[\frac{2  K}{r_{\textrm{thr}}} \, 
   \left(
1 - \frac{2\mathcal{A}_{0}}{E^{2}}
   \right) 
    - \frac{\mathcal{A}_{2}}{\mathcal{A}_{0}} 
   \right] \, u^{2}
    =  
    2 E^{2} \left( 1 - \sqrt{\frac{D^{2}}{D^{2}_{\text{crit}}} + \frac{2 \mathcal{A}_{0}}{E^{2}}} \right ) 
    \, ,
    \label{constraint3}
\end{equation}
where $D$ and $E$ are the impact parameter and energy of the timelike geodesic (respectively), and $D_{\text{crit}}$ is the critical impact parameter of the null geodesics. 
As seen in Fig.~\ref{fig:desv_tempo}, the approximation performed for timelike geodesics shows improved accuracy near the throat region, with decreasing concordance at larger distances.

The fixed points for both null and timelike geodesics are identified as the wormhole throat, which corresponds to the light sphere. Moreover, from results~\eqref{constraint2} and \eqref{constraint3}, we observe that the trajectories of these geodesics exhibit hyperbolic behavior close to the fixed point. Although the orbits have the same geometric shape, the trajectories of null and timelike geodesics do not coincide for the same values of energy $E$ and angular momentum $L$. This distinction is expected because the dynamical properties of null and timelike geodesics are inherently different. However, in the limit of large $E$, the trajectories of null and timelike geodesics converge. This behavior is consistent with relativistic principles, since timelike geodesics asymptotically approach null geodesics in the high-energy regime, reflecting the well-known phenomenon in which massive particles exhibit lightlike behavior as their velocities approach the speed of light.

Analogous to the massless particle result~\eqref{eq:sol1_hiper}--\eqref{eq:sol2_hiper}, result~\eqref{eq:linearizedSol} describes massive particles spiraling into or out of the wormhole throat. Massive particles stay near the throat for some time and eventually escape.

A qualitative difference between the dynamics of timelike geodesics and its counterpart for the null geodesics is the absence of a possible bifurcation in the present case. As can be seen from the expression~\eqref{potential}, the effective potential with $\mathscr{L}=1$ always has a quadratic form, even for radial geodesics ($L=0$). This excludes the kind of bifurcation observed in null-geodesic dynamics.

%%%%%%%%%%%%%%%%%%%%%%%%%%%%%%%%%%%%%%%%%%%%%%%%%%%%%%%%%%

\section{\label{sec:Final}Final remarks}

In this work, we investigated the dynamics of photons and massive particles around an asymptotically de Sitter wormhole. This spacetime was interpreted as a Randall-Sundrum brane. Geodesic stability was studied and the shadow of the wormhole was analyzed. We estimated the angular amplitude of the wormhole from the perspective of an observer near the throat, deriving analytical expressions within this approximation.

We found that the general behavior of null and timelike geodesics in the wormhole background is qualitatively similar. Both dynamics share a unique fixed point, corresponding to the wormhole throat, which is described as a photon sphere in our formalism. Stability evaluations using Lyapunov and Jacobi criteria consistently classify this structure as an unstable saddle point. Consequently, massless and massive particles are found to spiral into or out of the wormhole throat.

Bifurcation theory provided further insight into the null-geodesic dynamics. Notably, a Bogdanov-Takens bifurcation was identified, which manifests itself as a qualitative change in the effective potential for radial trajectories. It is interesting to note that this bifurcation does not appear in the dynamics of timelike geodesics. Moreover, the absence of homoclinic and heteroclinic trajectories for both null and timelike orbits suggests that these dynamical systems are structurally stable. These results contribute to a deeper understanding of wormhole physics, particularly regarding geodesic behavior and stability, and provide a foundation for future studies of observable phenomena.

\begin{acknowledgments}

R.~C.~B. acknowledges the partial support of the National Council for Scientific and Technological Development (CNPq), Brazil.
C.~M. is supported by Grant No.~2022/07534-0, S\~ao Paulo Research Foundation (FAPESP), Brazil.
\end{acknowledgments}

\section*{Data availability}

No data were created or analyzed in this study.

%%%%%%%%%%%%%%%%%%%%%%%%%%%%%%%%%%%%%%%%%%%%%%%%%%%%%%%

\end{document}